# A Fully Reconfigurable RF Vector Modulator-based Wideband Phase Shifter for NextG Beamforming Phased Array in Satellite Communications (SATCOM)

Hanxiang Zhang[1], Hao Yan[2], Hong Tang[3], Uzair Muhammad[2], Ayesha Naseem[2], Saeed Zolfaghary Pour[2], Po-Wei Liu[2], Fei Yan[2], Shehryar Niazi[2], Jintao Chen[2].

[1#] Department of Electrical Engineering and Computer Science, University of North Dakota, ND, USA
[2#] Florida A&M University - Florida State University College of Engineering, FL, USA
[3#] Department of Electrical & Computer Engineering, University of Massachusetts, Lowell, MA, USA

***Abstract*- This paper presents a fully reconfigurable RF vector modulator (RFVM)-based phase shifter for wideband beamforming phased array of 6G/NextG satellite communication (SATCOM). It covers the frequency range from *S*-band up to *Ku*-band. Specifically, the proposed RFVM features a novel vector modulation approach that relaxes the frequency constraint in conventional scenario while avoiding applying a quarter all-pass filtering (QAF) network. It achieves a topology that RF input signal is divided into multi-branch transversal sections where each branch enables individual gain weightings and phase delays. By re-summing the signals from all branches, the output signal of proposed RFVM can realize a full 360º phase shifting while it preserves magnitude response in wideband scenario.**

***Index Terms*- wideband phase shifting, beam-steering, sixth generation (6G), microwave circuits, transmission line, RF signal processing, vector modulation (VM), satellite communication.**

## I. Introduction

The mass deployment of non-geostationary satellites has been greatly promoting internet coverage and global connectivity, e.g. Kuiper, Starlink, OneWeb, and Iridium constellation. Those low earth orbit (LEO) satellite systems drastically enhance wireless communication by providing high data capacity, low transmission latency, increased power efficiency, etc. In this era, beamforming phased array plays an essential role due to its capability of electronical beam-steering to establish the optimum wireless transmission links [1]-[2]. It can realize the flexible spatial division multiplexing access (SDMA) to increase spectral efficiency and system capacity. The conventional phased arrays are typically optimized for single-band operation and only cover a portion of the available spectrum in satellite communications (SATCOM), such as *Ku*-, *K*-, and *Ka*-bands [2]-[6]. However, beyond the 5G FR-2 that bandwidth emerged with 2:1 bandwidth at 24GHz-40GHz and later expanded to 15GHz-55GHz, the 6G FR-3 operating from 7.125GHz to 24.25GHz poses a significant challenge that phased arrays are required to work well in a wideband scenario [7]-[8].

As the vital module, phase shifters create progressive phases on RF signals feeding on the antenna array directly contributes to the spatial radiation beam-steering. Among prior works, a lot of types of phase shifter designs have been investigated. For example, switched-line/loaded-line phase shifters obtain phase shifts by switching between transmission lines of different electrical lengths and periodically loading a transmission line with reactive elements [9]-[11]. Feeding matrix networks create phase shift through frequency-dependent phase response while maintaining constant magnitude [12]-[14]. However, they usually own moderate phase tuning range, coarse phase step resolution, and limited operating bandwidth. To advance the progressive phase characteristics, the true time delay (TTD) modules, in-phase and quadrature (I-Q) vector modulators are rapidly rising.

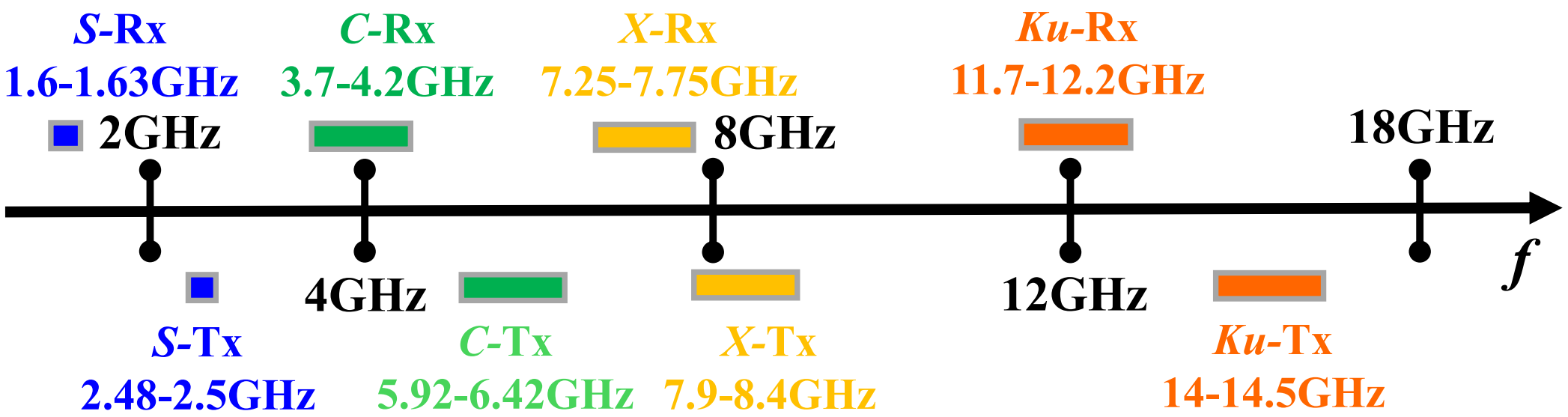

Fig. 1 Spectrum distributions of the satellite communication with up-link and down-link from *S*-band to *Ku*-band.

Specifically, TTD module applies constant time delays to input signals, leading to the corresponding phase shift linearly scaled with frequency, such that it effectively conquers beam-squinting by preserving beam directions across wide bandwidth [15]-[16]. However, several trade-offs are necessary between its high insertion loss, discrete delay resolution, and complex calibration [17]-[18], etc. On the other hand, I-Q modulator has been attracting great interest that benefits from its continuous phase control, low switching artifacts, and high reconfigurability [19]. Conventionally, the input signal is decomposed into four orthogonal in-phase and quadrature components (±I, and ±Q) with a 90º relative phase differences through a quadrature all-pass filtering (QAF) network. By independently weighing and recombining these vectors, an arbitrary phase (and magnitude) can be synthesized accordingly. In the past decades, I-Q modulator-based phase shifting has been intensively implementing in the phased array antenna system and making big achievements. For example, [20] proposes the beamformer flip-chip cell for a 32-element phased array transceiver (TXCVR), and [21] presents a design of programmable phased array receiver (Rx) with different simultaneous beams generation.

Besides, the wideband phased arrays applications have been also widely investigated through I-Q modulation approach. For examples, two types of reconfigurable wideband phased array Rxs were presented for 6G FR3 system [22]-[23]. In [24], an 8-channel wideband phased array transmitter (Tx) is introduced for SATCOM applications. Additionally, several other wideband phased arrays were explored to operate across different spectrum bands [25]-[28]. All those works have proved the capability of I-Q modulation-based wideband phased array that relying on vector synthesis of orthogonal signal componentss to achieve agile, continuous beam-steering with integratioon advantages. However, the performance of I-Q modulator is fundamentally constrained by how closely the circuit can approximate its ideal assumptions. The accurate phase shifting often requires precise 90º relative phase division with minimum amplitude imbalance between I- and Q- paths over the operating frequencies. In practice, the conventional quadrature-generation components, such as hybrid couplers [29]-[30] and QAFs [31]-[32], exhibit a finite bandwidth, frequency-dependent imbalance, and excessive losses. Also, high linearity in both I- and Q- paths is required to prevent vector distortions and phase errors across a wide bandwidth.

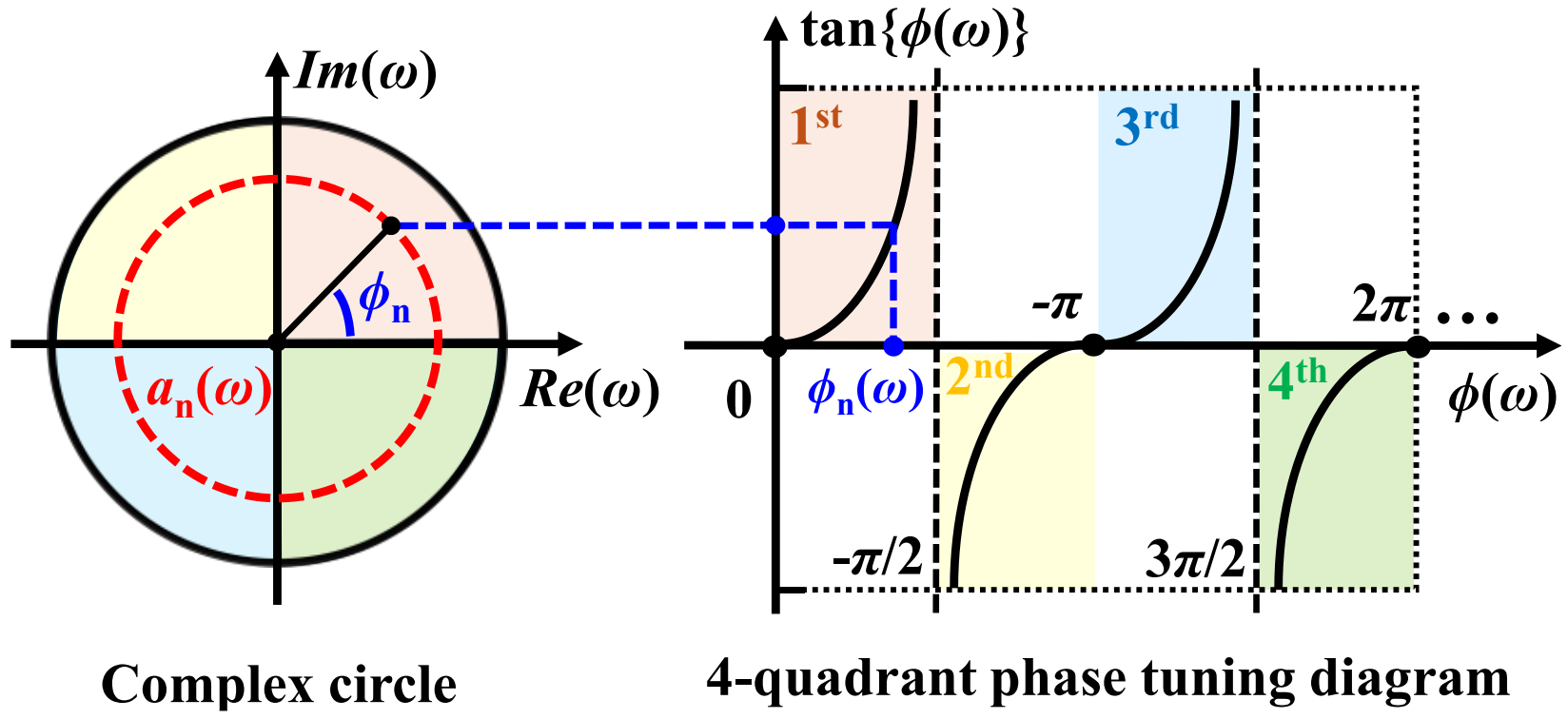

Fig. 2 Four-quadrant phase shifting that described in a complex plane.

Inspired by the prior works and related challenges, a novel reconfigurable RF vector modulator (RFVM)-based wideband phased array has been proposed and designed for SATCOM from *S*-band up to *Ku*-band, as in Fig. 1. Different from other designs, the proposed design applies a transversal multi-branch FIR network topology to conduct the real-time wideband RF signal processing. It implements the differential conversion on the input signal to relax the degradation of conventional polyphase network (qudrature transformation), while it achieves more flexibility on magnitude/phase responses within individual spectrum sub-bands.

## II. Design of the Proposed RF Vector Modulator-Based Phase Shifter

### A. Analysis of Phase Shifter Transfer Function

For a RF phase shifter design, its general transfer function is expressed as:

$$H(\omega) = a(\omega) \cdot e^{j\phi(\omega)} \tag{1}$$

Where $0<\omega<\omega_0$ is the desired operating bandwidth, $a(\omega)$ and $\phi(\omega)$ denote its magnitude and phase responses, respectively. In this representation, the phase response is defined over full angular range of $2\pi$, such that the spectral response is naturally interpreted as a rotating vector in the complex plane, as illustrated in Fig. 2. Accordingly, the phase state of a phase shifter is mapped onto the four quadrants of a complex plane through a value of $\phi(\omega)$.

To realize the complete four-quadrant phase tuning over the selected band, two generalized phase parameters are introduced in the phase components as:

$$\phi_q = \frac{\pi}{2}(a+\gamma) \tag{2}$$

where $\alpha \in [-1,1]$ is a continuous phase-tuning factor and $\gamma = [-1,1]$ is a discrete angular-polarity selector. Under this definition, $\alpha$ governs the continuous phase variation, whereas $\gamma$ determines the corresponding phase-offset region. More specifically, when $\gamma = \pm 1$, the resulting phase spans the interval $[-\pi, 0)$ and $(0, \pi]$, respectively. The combined use of $\alpha$ and $\gamma$ enables continuous $2\pi$-phase coverage over the selected operating band. Accordingly, the generalized transfer function that realizing quadrant selectable phase shifting is derived as:

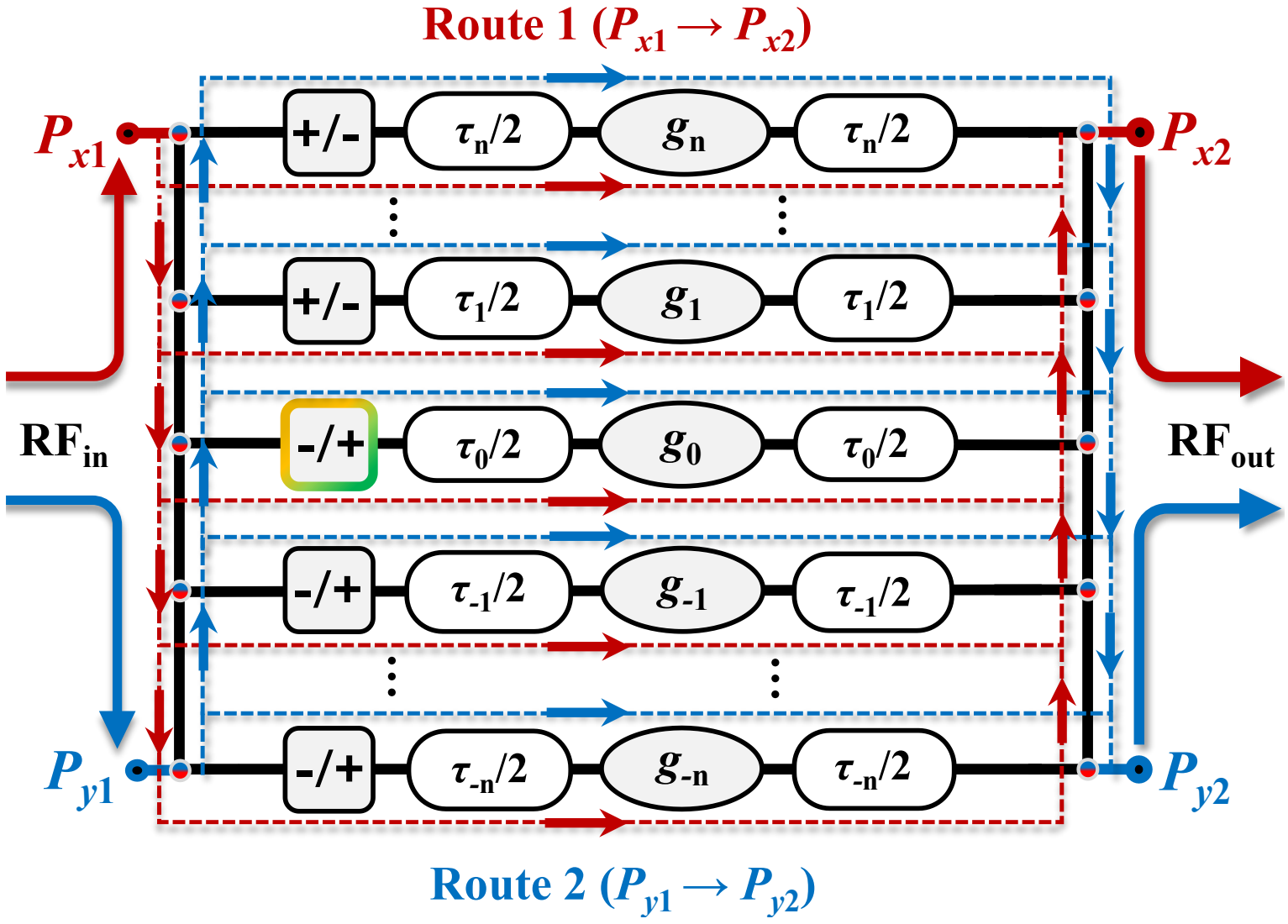

Fig. 3 Conceptual diagram of the proposed reconfigurable RF vector modulator phase shifter in topology of a transversal FIR network.

$$H_{\alpha,\gamma}(\omega) = a(\omega) \cdot e^{j \cdot \phi_q} = a(\omega) \cdot e^{j \cdot \frac{\pi}{2}(\alpha+\gamma)}, (0 < \omega < \omega_0) \tag{3}$$

To relate the target response in (3) to a realizable time-domain form, the corresponding impulse response is obtained through inverse Fourier transformation over the desired bandwidth, yielding:

$$h_{\alpha,\gamma}(t) = \frac{1}{2\pi} \int_0^{\omega_0} H_{\alpha,\gamma}(\omega) \cdot e^{j\omega t} d\omega \tag{4}$$

Substituting (4) into (3), it then gives:

$$h_{\alpha,\gamma}(t) = e^{j\frac{\pi}{2}(\alpha+\gamma)} \cdot \frac{1}{2\pi} \int_0^{\omega_0} a(\omega) \cdot e^{j\omega t} d\omega \tag{5}$$

Equation (5) shows that the desired phase state contributes a common complex rotation, whereas the kernel shape is determined by the magnitude response of $a(\omega)$. For practical implementation, the continuous-time kernel in (5) must be represented by a finite set of delayed and weighted signal components. Evaluating the kernel at prescribed branch delays $\tau_n$ yields the corresponding effective coefficients:

$$c_n(\alpha,\gamma) \approx h_{\alpha,\gamma}(\tau_n) = e^{j\frac{\pi}{2}(\alpha+\gamma)} \cdot \frac{1}{2\pi} \int_0^{\omega_0} a(\omega) \cdot e^{j\omega\tau_n} d\omega \tag{6}$$

Equations (4) - (6) reveal that the branch-gain envelope is governed by the Fourier projection of the target magnitude response over the operating band, whereas the parameters $\alpha$ and $\gamma$ determine the effective phase state. Therefore, the desired phase-shifting response with preserved magnitude can be synthesized through a finite set of delayed and weighted signal replicas, which naturally leads to a transversal finite impulse response (FIR) network topology. The corresponding topology and practical coefficient distribution are introduced in the following section.

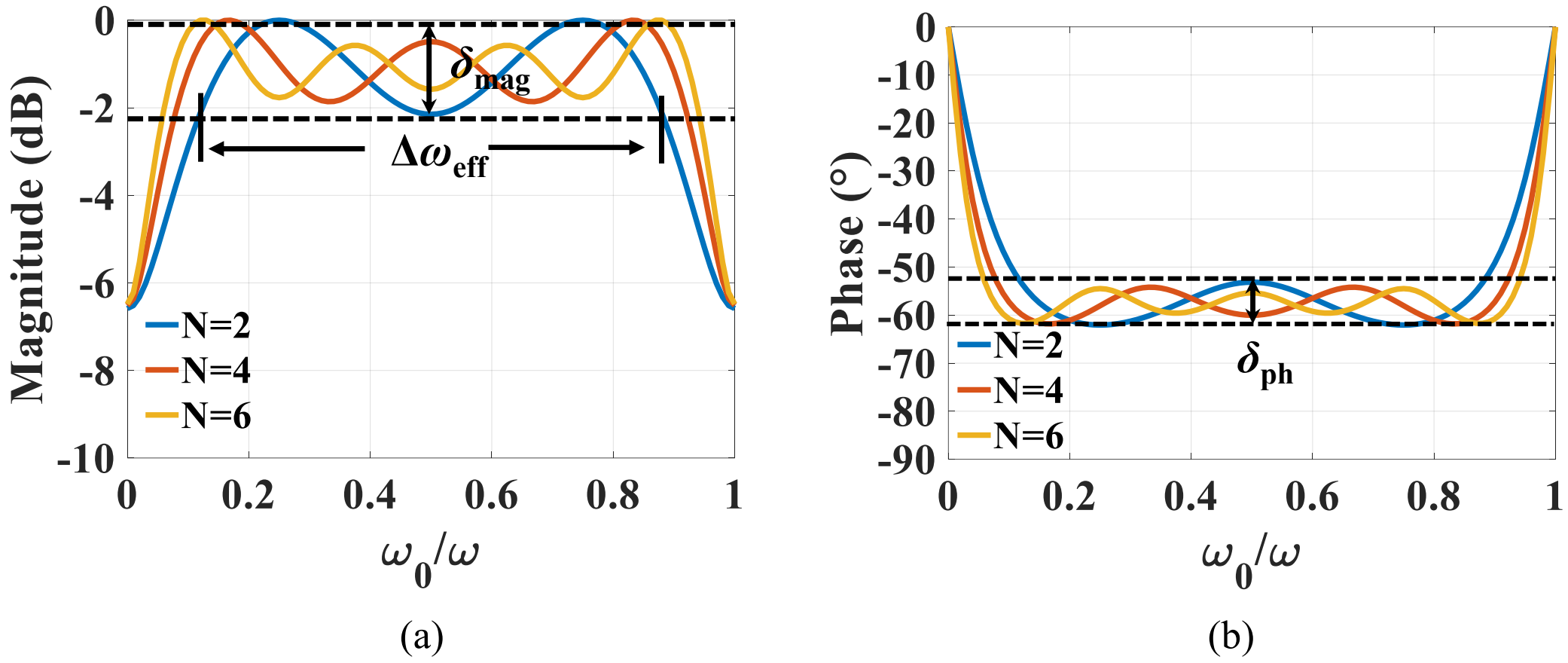

Fig. 4 Transfer function descriptions of the proposed RFVM with varying transversal tap number $N$ = 2,4,6 (a) magnitude response (b) phase response.

*B. Synthesis of the Proposed Reconfigurable RFVM*

Fig. 3 shows a schematic diagram of the proposed RF vector modulator. As illustrated above, it is in form of a weighed multi-delay network. Different from the conventional single-chain FIR representation, the proposed RFVM is organized as a dual-route reconfigurable signal-propagation network. In this topology, the RF input signal is split into two complementary propagating routes, denoted as Route 1 ($P_{x1}$-to-$P_{x2}$) and Route 2 ($P_{y1}$-to-$P_{y2}$), respectively. These two routes traverse through the same set of weighted delay branches and are subsequently recombined at the output side. Here, the impulse response and its corresponding transfer function of a (2$N$+1)-branch transversal network are given as:

$$h_{\alpha,\gamma}(t) \approx \sum_{n=-N}^{N} g_n(\alpha,\gamma)\cdot\delta[t-(-1)^n\cdot(2|n|-1)T] \quad (7)$$

$$H_{\alpha,\gamma}(\omega) \approx \sum_{n=-N}^{N} g_n(\alpha,\gamma)\cdot e^{(-1)^n\cdot j\omega\cdot(2|n|-1)T},\ 0<\omega<\omega_0 \quad (8)$$

where $T$ is the elemental spacing and $g_n(\alpha, \gamma) = |c_n(\alpha, \gamma)|$ denotes the magnitude of effective coefficients for the $n$-th branch. In this topology, each horizontal branch contributes an individual weighted and delayed component to the synthesized response. More specifically, the center branch ($g_0$, $\tau_0$) serves as reference branch, whereas the remaining upper ($g_n$, $\tau_n$) and lower branches ($g_{-n}$, $\tau_{-n}$) correspond to the delayed terms that indexed by positive and negative orders relative to the center branch. It is noticed that the total delay is symmetrically distributed as $|\tau_n/2|$ before and after weighing blocks in each branch. For the proposed RFVM, the branch-gain distribution is prescribed to follow the odd-harmonic coefficient law that is inspired by discrete Fourier series expansion. Accordingly, the normalized gain envelope is chosen as:

$$g_n = \begin{cases} g_0, & n=0 \\ \dfrac{1}{2|n|-1}, & n\neq 0 \end{cases} \quad (9)$$

To incorporate the reconfigurable phase control introduced in section II-A, the actual signed coefficients are assigned through the control factors $\alpha$ and $\gamma$. Then, the equation (9) can be further derived as:

$$g_n(\alpha,\gamma) = \begin{cases} \gamma, & n = 0 \\ \mathrm{sgn}(n)\cdot\dfrac{a\cdot\gamma}{(2|n|-1)}, & n \neq 0 \end{cases} \tag{10}$$

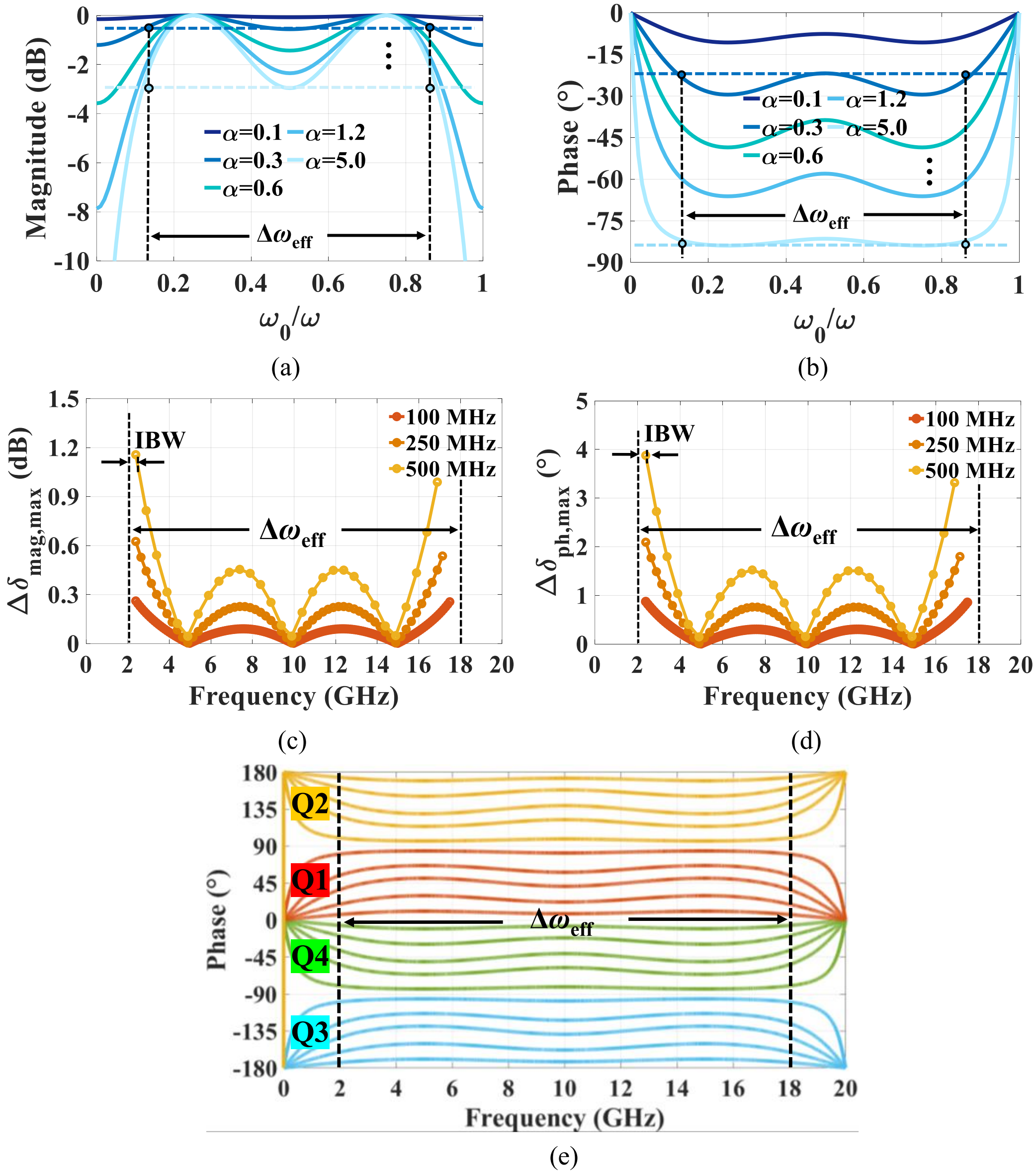

Fig. 5 Analysis diagrams of the proposed real-time RFVM to realize a 4-quadrant continuous phase shifting (a) magnitude response (b) relative phase response (c) maximum $\Delta\delta_{mag}$ ($\alpha = 5.0$) for different IF-bands (d) maximum $\Delta\delta_{ph}$ ($\alpha = 0.6$) for different IF-bands (e) 4-quadrant (full 360º) phase shifting from 2GHz to 18GHz.

Equation (10) directly reveals that four selectable coefficient modes correspond to the combinations of signed ratio $\alpha$ and polarity selector $\gamma = [-1,1]$, which enable a continuous four-quadrant phase-state control. Therefore, the transfer function can be explicitly expressed as:

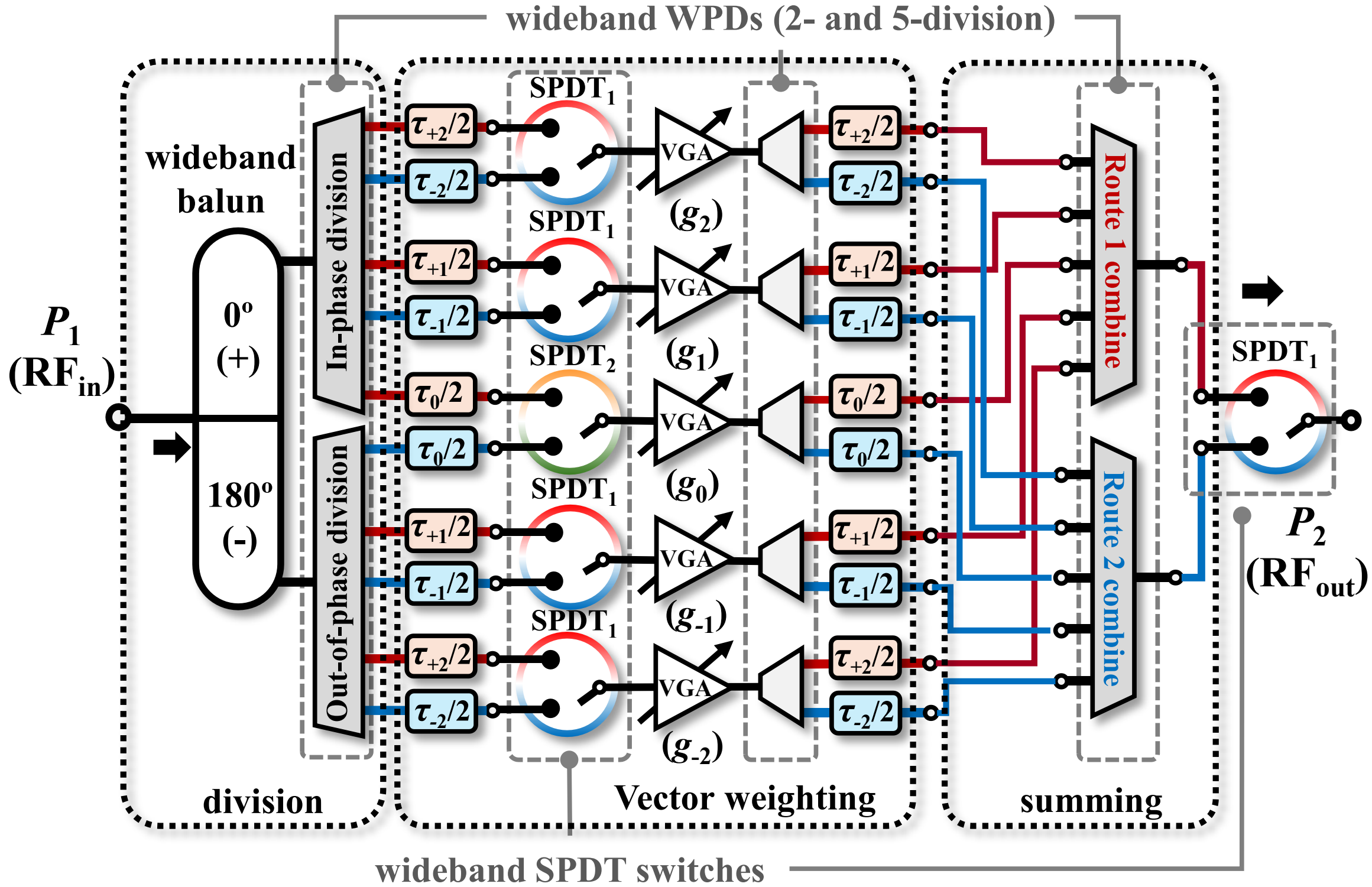

Fig. 6 Circuit implementation of the proposed reconfigurable RFVM phase shifter with 5 transversal branches topology, it consists of three functional modules of signal division, signal modulation, and signal re-combination.

Table I four-quadrant phase mode selection of RFVM and its corresponding circuit configuration

| Mode | Phase Range | Circuit Configuration | | | Parameter Configuration | |
|---|---|---|---|---|---|---|
| | | Route | SPDT$_1$ | SPDT$_2$ | $\alpha$ polarity | $\gamma$ polarity |
| Q1 | 0° ~ +90° | 1 | up | up | + | + |
| Q2 | +90° ~ +180° | 2 | up | down | + | - |
| Q3 | -180° ~ -90° | 2 | down | down | - | - |
| Q4 | -90° ~ 0° | 1 | down | up | - | + |

$$H_{\alpha,\gamma}(\omega)=\gamma\cdot\left[g_0+\alpha\sum_{n=-N}^{N}g_n(\alpha,\gamma)\cdot e^{(-1)^n\cdot(2|n|-1)T}\right] \tag{11}$$

Here, Fig. 4 shows the frequency responses of proposed RFVM by letting parameters $\alpha=\gamma=1$ and varying the tap pair number $N$. In Fig. 4 (a), the magnitude response shows a truncation passband across the normalized frequency that owning in-band root mean square (RMS) deviation of $\delta_{\text{mag}}$. The curves appear the ringing effect within the operating band (Gibbs phenomenon), which naturally reveals a synthesis of finite number of Fourier series. When increasing the tap pair number of $N$ = 2, 4, 6 (or even larger), the response curve is approaching an ideal square wave with smaller $\delta_{\text{mag}}$ and wider effective bandwidth $\Delta\omega_{\text{eff}}$ (limited by attenuation of max$\{|\delta_{\text{mag}}|\}$). Similarly, in Fig. 4 (b), the phase response shows a square truncation passband curve with phase deviation $\delta_{\text{ph}}$, which can also be minimized with larger tap number. Therefore, 4-quadrant phase shifting can be decently realized by configuring the value of factors $\alpha$, $\gamma$, and the analysis diagrams are explicitly described in Fig. 5.

Fig. 5 (a) - (b) illustrates the four-quadrant continuous phase shifting realization of a 5-tap RFVM ($N = 2$) across the desired bandwidth based on equations (10) - (11), where the normalized gain envelope distribution is accordingly taken as $|g_n|$ = 1, 1/3, 1/5. By setting $\gamma = 1$ and appropriately sweeping ratio factor $\alpha$ = 0.1, 0.3, 0.6, 1.2, 5.0, the phase of RFVM can be shifted up to 86º with corresponding magnitude responses. To be noticed, within the effective bandwidth of $\Delta\omega \approx 0.8 \cdot \omega_0$, the maximum deviations of in-band magnitude and phase are shown as $\Delta\delta_{mag}$ = 3dB ($\alpha = 5$) and $\Delta\delta_{ph}$ = 9.8º ($\alpha = 0.6$). However, those deviations are becoming negligible for the large value of $\omega_0$ within a relatively small instantaneous bandwidth (IBW).

As illustrated in Fig. 5 (c) - (d), for $f_0 = 2\pi\omega_0$ = 20 GHz (effective operating frequency ranges from 2 GHz to 18 GHz), $\Delta\delta_{mag}$ and $\Delta\delta_{ph}$ are limited to 1.18 dB and 3.9º respectively when its IBW is smaller than 500MHz (the typical IBW in SATCOM spectrum). Therefore, a full 360º phase shifting can be realized and continuously tuned by appropriately configuring the values of $\alpha$ and $\gamma$. In Fig. 5 (e), the modes of four-quadrant phase can be switched and shifted between -180º and +180º (Q1: 0º ~ 90º, Q2: +90º ~ +180º, Q3: -180º ~ -90º, Q4: -90º ~ 0º) with decent magnitude responses, as shown in Fig. 5 (e).

## III. Conclusion

In this paper, a novel real-time RF vector modulator-based phase shifter has been presented. It can be reconfigured to achieve full 360º phase shifting with preserving magnitudes. Different from the conventional vector modulation approaches using a quadrature all-pass filter network plus I-Q modulator, the proposed RFVM utilizes a differential signal conversion to achieve four-quadrant phase mode. In the future work, the proposed RFVM shall be designed and integrated into wideband beamforming transceiver link for further measurements and system level over-the-air (OTA) evaluations.